\documentclass[prl,showpacs,superscriptaddress,twocolumn]{revtex4}
\usepackage{amsmath}
\usepackage{amsfonts}
\usepackage{amssymb}
\usepackage{graphicx}
\usepackage{color}
\definecolor{MyDarkGreen}{rgb}{0,0.5,0.0}
\def\tj{t_{jam}}

\def\pa{\phi_{J_1}}
\def\pb{\phi_{J_2}}
\def\pc{\phi_{J_3}}

\begin{document}
\title{Jamming phase diagram for frictional particles}
\author{Massimo Pica Ciamarra}
\affiliation{Department of Physical Sciences, University of Naples `Federico II', Coherentia-CNR, Napoli, Italy}
\author{Raffaele Pastore}
\affiliation{Department of Physical Sciences, University of Naples `Federico II', Coherentia-CNR, Napoli, Italy}
\author{Mario Nicodemi} 
\affiliation{Complexity Science \& Department of Physics, University of Warwick, UK}
\author{Antonio Coniglio}
\affiliation{Department of Physical Sciences, University of Naples `Federico II', Coherentia-CNR, Napoli, Italy}
\affiliation{INFN, Sezione di Napoli, Italy}
\begin{abstract}
The non-equilibrium transition from a fluid--like state to a disordered solid--like state, known as the jamming transition, occurs in a wide variety of physical systems, such as colloidal suspensions and molecular fluids, when the temperature is lowered or the density increased. 
Shear stress, as temperature, favors the fluid-like state, and must be also considered to define the system `jamming phase diagram' \cite{Liu98,Trappe01,Ohern03,Coniglio04}. Frictionless athermal systems\cite{Liu98}, for instance, can be described by the zero temperature plane of the jamming diagram in the temperature, density, stress space. 
Here we consider the jamming of athermal frictional systems\cite{Cates98,Makse2000,Makse2005,grebenkov,Saarloos2007,Makse2008} such as granular materials, which are important to a number of applications from geophysics to industry. At constant volume and applied shear stress\cite{Liu98,Trappe01}, we show that while in absence of friction a system is either fluid-like or jammed, in the presence of friction a new region in the density shear--stress plane appears, where new dynamical regimes are found. In this region a system may slip, or even flow with a steady velocity for a long time in response to an applied stress, but then eventually jams. 
Jamming in non-thermal frictional systems is described here by a phase diagram in the density, shear--stress and friction space. 
\end{abstract}

\maketitle

Our analysis is based on Molecular Dynamics simulations of a suspension of soft-core spherical grains\cite{Silbert} enclosed between two rough plates at constant volume fraction $\phi$, and subject to a constant shear stress $\sigma$. Periodic boundary conditions are used in the other directions (detail on the system and on the numerical method are given in the Supplementary Information). When the Coulomb friction coefficient $\mu$ is set to zero, this model reduces to an assembly of frictionless particles, which, at zero applied stress $\sigma = 0$, jam at the random close packing volume fraction\cite{Ohern03} $\phi_J(\sigma = 0) \simeq 0.64$. 
For $\phi > \phi_J(0)$ the shear and the bulk modulus grow as $\phi$ increases. When the system is subject to a small applied shear stress $\sigma$, it flows when $\phi$ is less than the jamming threshold $\phi_J(\sigma)$, while it responds as an elastic solid when $\phi > \phi_J(\sigma)$.

\begin{figure}
\begin{center}
\includegraphics*[scale=0.27]{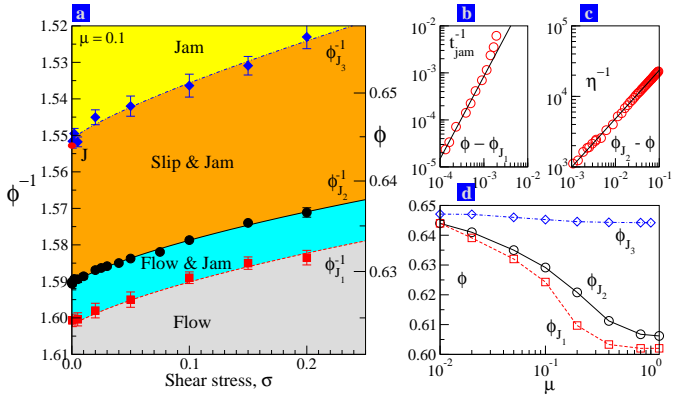}
\end{center}
\caption{\label{fig:diagram}
{\bf Flowing regimes of a granular system under shear stress.}  
{\bf a} In the ``Flow'' region the system flows with a steady velocity;  in the ``Flow \& Jam'' region the system first flows with a steady velocity, but jams after a time $\tj$;  in the ``Slip \& Jam'' region the system slips of a distance $\Delta L$, never reaching a steady velocity, and then jams.  In the ``Jam'' region the system deforms as a solid as soon as the shear stress is applied. The jamming time $\tj$, the viscosity $\eta$ and the slip distance $\Delta L$ have been fitted by power laws, $\tj \sim (\phi-\pa)^{-\alpha}$, $\eta \sim (\pb - \phi)^{-\gamma}$, $\Delta L \sim (\pc-\phi)^{\beta}$ for any given value of $\sigma$ and $\mu$. The volume fractions at which $\tj$ diverges, $\eta$ diverges, and $\Delta L$ vanishes are shown, respectively, as squares ({\color{red}{$\blacksquare$}}), circles ({\color{black}{\textbullet}}) and diamonds ({\color{blue}{$\blacklozenge$}}). 
For $\sigma = 2~10^{-3}$ and $\mu = 0.1$, ${\bf b}$ shows the power law divergence of $\tj$ in ${\phi-\pa}$ with $\alpha = 1.75$ at $\pa = 0.622$, and ${\bf c}$ that of the viscosity in $\pb-\phi$ with $\gamma = 0.75$ at $\pb = 0.625$. {\bf d} Dependence of $\pa$, $\pb$ and $\pc$ on $\mu$ at $\sigma = 2~10^{-3}$. They conincide at $\mu = 0$ and decreases as $\mu$ increases.
}
\end{figure}

At $\sigma = 0$, the presence of friction is known to modify the location of the jamming point, which becomes also dependent on the preparation protocol\cite{Makse2000,Makse2005,Saarloos2007,Makse2008}. Here we focus on $\sigma > 0$, and show that friction gives rise to new dynamical regimes as  illustrated in the phase diagram of Figure~\ref{fig:diagram}a, where the flowing properties of a granular system with friction coefficient $\mu = 0.1$ are summarized in the inverse density, $\phi^{-1}$, and shear stress, $\sigma$, plane. The initial state is prepared in such a way that no frictional contacts are present\cite{Makse08}, a condition which can be experimentally realized via high-frequency small amplitude vibrations\cite{Ohern08}. 
At low density, in the `Flow' regime, the system flows and reaches a stationary velocity. For $\phi$ larger than a threshold $\pa = \pa(\sigma,\mu)$, the system enters the `Flow \& Jam' region. Here the system first flows with a stationary velocity (reached after a transient), but eventually enters by chance a microscopic configuration which is able to sustain the applied shear stress, and jams. The `Flow \& Jam' region is limited by a jamming line $\pb = \pb(\sigma,\mu)$. Above $\pb$ steady flow is never observed, and the system jams after a small slip. This  `Slip \& Jam' region is limited by the line $\pc(\sigma,\mu)$ above which the system does not slip, but responds as a solid to an applied external stress. Examples of the system time-course in these different regions are illustrated in Fig. S2.

How to define quantitatively the three lines? In the `Flow \& Jam' region the system stops flowing after a jamming configuration has been selected. The average time $\tj$ the system flows before jamming is longer the smaller the volume fraction, and diverges approaching the `Flow' regime, where it is actually infinite. $\pa$ is defined as the volume fraction where $\tj$ diverges coming from above (see Fig.~\ref{fig:diagram}b).
The line $\pb$ is defined by the divergence of the shear viscosity $\eta(\phi,\sigma,\mu)$, defined as the ratio between shear stress $\sigma$ and shear rate $v_{s}/h$,  where $h$ is the distance between the two plates, and $v_{s}(\phi,\sigma,\mu)$ the shear velocity\cite{note}. The viscosity increases as a power law with the volume fraction $\phi$, and diverges as $\phi$ approaches $\pb$, i.e. $\eta \propto (\pb-\phi)^{-\gamma}$ (see Fig.~\ref{fig:diagram}c). The exponent $\gamma$ appears not to depend on the shear stress, while it depends on the friction coefficient.
The line $\pc$ marks the end of the `Slip \& Jam' region, as detailed in the Supplementary Information. 
We have defined the slip as the residual displacement of the top plate in a stress cycle, which allows to separate the slip of the top plate from its displacement due to the elastic deformation of the system on jamming.
After preparing the system, we slowly increase the stress to its final value $\sigma$, and then decrease it to zero. Below $\pc$, this cycle is irreversible and the initial and final position of the top plate differ by the slip distance $\Delta L$, while above $\pc$ the cycle is reversible and $\Delta L = 0$. 

As discussed in the Supplementary Information, the line $\pa$ and $\pb$ of the the diagaram of Fig.~\ref{fig:diagram} do not depend on the preparation protocol, for instance if friction is considered when inflating the particles in the preparation of the initial state, as in several studies\cite{Makse2005,Saarloos2007}. This is because these lines are determined from extrapolations in the flowing state where correlations with the initial state are by definition lost. The line $\pc$ may depend on the protocol. The one we have determined is an upper bound with respect to those generated by all other protocols.

$\pa$, $\pb$ and $\pc$ depend on the friction coefficient $\mu$. Fig.~\ref{fig:diagram}d illustrates their dependence on $\mu$ for $\sigma = 2~10^{-3}$. The three lines coincide at $\mu = 0$, if possible dynamical effects are neglected\cite{PCC}, but decreases with $\mu$ in such a way that $\pc(\sigma,\mu) > \pb(\sigma,\mu) > \pa(\sigma,\mu)$.  In the limit of high friction, each line reaches a plateau. The dependence of $\pc$ on $\mu$ is very small, and only appears at high $\phi$ or $\sigma$, due to the presence of a plastic response of the system (see Fig. S4).

\begin{figure}
\begin{center}
\includegraphics*[scale=0.5]{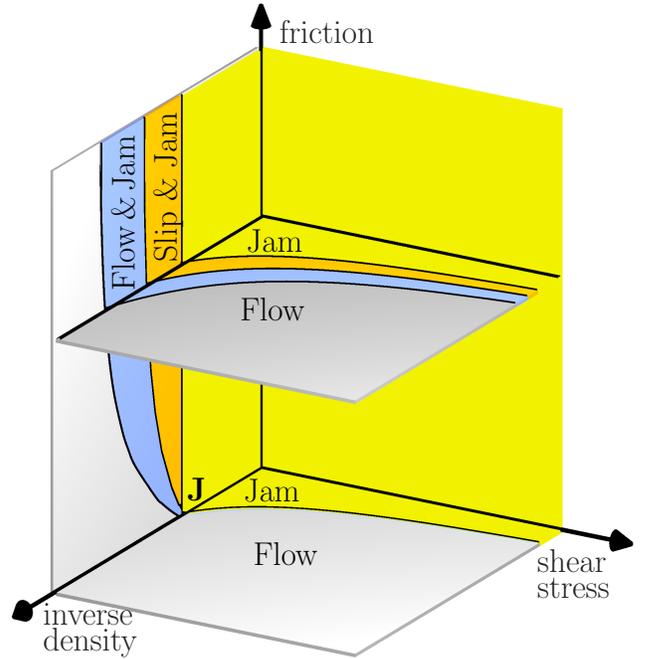}
\end{center}
\caption{\label{fig:3ddia}
{\bf Jamming phase diagram for frictional system.}
The jamming properties of frictional systems are illustrated in a diagram with axis the inverse density, the shear stress and the friction coefficient. At zero friction the jamming phase diagram is characterized by a ``Flow'' and by a ``Jam'' region, while in the presence of friction two new regions appear: the ``Flow \& Jam'' region and the ``Slip \& Jam'' region.
}
\end{figure}
Such a dependence on friction leads to a jamming phase diagram for frictional particles characterized by three axis: the inverse density, the shear stress and the friction coefficient. In this phase diagram, schematically shown in Fig.~\ref{fig:3ddia},
the surfaces $\pa(\sigma,\mu)$, $\pb(\sigma,\mu)$ and $\pc(\sigma,\mu)$ enclose the regions of different flow properties.
The surfaces collapse in a line in the zero-friction plane, which coincides with the zero-temperature plane of the jamming phase diagram of frictionless particles.
The phase diagram of Fig.~\ref{fig:3ddia} clarifies the intuitive expectation that, when a frictional system jams after flowing, then it is possible to unjam it not only varying the density or the shear stress, but also by changing the friction coefficient (which depends on humidity, temperature as well as on the presence of lubricants\cite{Yoshizawa,Crassous}).
The smallest value of the density at which we found jammed states is $\pa(\sigma \to 0) \simeq 0.585$, close the smallest value reported in the litterture for jammed states in absence of gravity\cite{Makse2005}. In the presence of gravity, looser states have been found\cite{Onoda}.

\begin{figure}
\begin{center}
\includegraphics[scale=0.33]{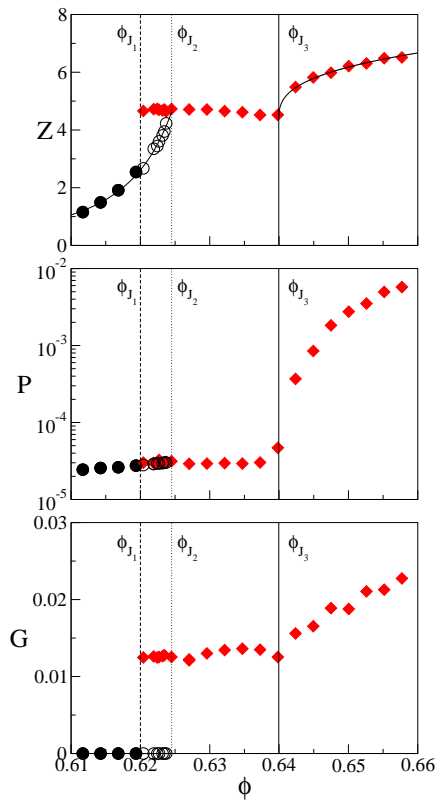}
\end{center}
\caption{\label{fig:dinamica}
{\bf Mechanical properties} Plot of the mean contact number $Z$ ({\bf a}), of the normal pressure on the shearing plate $P$ ({\bf b}), and of the shear modulus $G$ ({\bf c}) as a function of $\phi$, for $\sigma = 2~10^{-3}$ and $\mu = 0.1$.  The vertical lines mark $\pa$ (dashed), $\pb$ (dotted), and $\pc$ (plain). Circles are measures taken when the system flows, while diamonds are measure taken in jammed configurations. Open circles in the range $\pa$--$\pb$ are measures taken in the flowing regime for $t < \tj(\phi)$, before the system jams.}
\end{figure}

The changes in the structural properties of the system which occur crossing $\pa$, $\pb$ and $\pc$ are shown 
in Fig.~\ref{fig:dinamica} for $\mu = 0.1$ and $\sigma = 2~10^{-3}$. We discuss here the volume fraction dependence of the mean contact number $Z$, of the normal pressure on the confining walls $P$, and of the shear modulus $G$. The measure of $G$,  which is discussed in the Supporting Information, is only possible because the system, despite having and anysotropic microstructure, is not fragile\cite{Cates98} and responds elastically and almost isotropically to small external perturbations.
In the flowing regime (circles) $Z$ and $P$ increase with $\phi$, while $G$ is zero. In the jammed regime (diamonds) $Z$, $P$ and $G$ are roughly constant for $\phi < \pc$, while they increase as power laws for $\phi > \pc$, where a continuous transition occurs. Measures taken in the flowing state in the range $\pa$--$\pb$ (open circles) are taken for $t < \tj(\phi)$, before the system jams. These results are in qualitative agreement with recent experimental results\cite{PrlBehringer} which have also show that, contrary to the frictionless case,  $Z$ and $P$ do not vanish at the random close packing volume fraction in the presence of friction. Compared to previous numerical studies\cite{Makse08,Makse2005,Saarloos2007} conducted at $\sigma = 0$, our findings clarify that there is a whole volume fraction range where frictional granular systems may have the same mechanical properties. This volume fraction range can be identified with a constant $Z$ line of the recently introduced $Z$-$\phi$ diagram\cite{Makse08}.

We have shown here that friction strongly influences the jamming properties of particulate systems. An open question is to explain the presence of the flow and jam region, where flowing frictional systems subject to a constant shear stress suddenly jam. In constant volume systems, it is known that when the shear rate is fixed, large normal stresses fluctuations occur\cite{fluctuationstresses0,fluctuationstresses1,fluctuationstresses2}, as a consequence of a frustrated dialtancy\cite{Reynolds}.
When the shear stress is fixed, as in the case considered here, strong fluctuations are not observed in the normal stresses, but in the shear rate. We speculate that at constant $\sigma$ a system jams in correspondence of a fluctuation so large that the shear velocity vanishes (see Supplementary Informations). Therefore, the flow and jam phenomenology at constant shear stress can be seen as the counterpart of the large normal stress fluctuations observed at a constant shear rate\cite{fluctuationstresses0,fluctuationstresses1,fluctuationstresses2}.
As jamming occurs when a percolating cluster of particles builds up, this argument may be related to the k--core percolation model introduced to describe jamming of frictionless particles\cite{Schwarz}, or to models which explicitely take into account the constraint of mechanical equilibrium on each grain\cite{Lois,Bulb}. However, it must be considerd that while in absence of friction the percolating cluster can only emerge due to changes in the control parametrs, in the presence of friction such a cluster may spontaneously emerge while the system is flowing. 
The role of temperature in the jamming of frictional particles should be also investigated\cite{Liu09}. Large colloidal particles, with a size smaller than roughly $1\mu m$, are in fact at the same time small enough for temperature to influence their dynamics, and large enough to be characterized by frictional forces. Dense colloidal suspensions have actually already shown to behave as dense granular systems\cite{Ballesta}.

\setlength{\oddsidemargin}{0.2cm}
\setlength{\textwidth}{15cm}
\begin{widetext}
\newpage

\def\tj{t_{\rm jam}}
\def\fs{{\it fragile states} }

\def\pa{\phi_{J_1}}
\def\pb{\phi_{J_2}}
\def\pc{\phi_{J_3}}


\thispagestyle{empty}
\hrule ~\\
\begin{center}
{\bf \Large Supplementary Informations}\\
\end{center}
\begin{center}
{\bf \large Jamming phase diagram for frictional particles}\\
\end{center}
\begin{center}
{Massimo Pica Ciamarra, Raffaele Pastore, Mario Nicodemi, Antonio Coniglio}
\end{center}

\tableofcontents

\newpage

\section{Numerical model\label{Sec:numericalmodel}}
\subsection*{Interaction between grains}
We have performed Molecular Dynamics simulations of a monodisperse system of particles of mass $M$ and diameter $D$, based on a standard model for the grain-grain interaction, which is the linear spring-dashpot model. 
Two particles $i$ and $j$, in positions ${\bf r}_i$ and ${\bf r}_j$, with linear velocities ${\bf v}_i$ and ${\bf v}_j$, and angular velocities ${\bf \omega}_i$ and ${\bf \omega}_j$, interact if in contact.The interaction force has a normal and a tangential component. \\
\noindent The normal component is given by:
\[
 {\bf F_{n_{ij}}} = -k_n \delta_{ij} {\bf n}_{ij} - \gamma_n m_{eff} {\bf v}_{{\bf n}_{ij}},
\]
where $k_n$ is the elastic modulus of the particles, ${\bf r}_{ij} = {\bf r}_{i} - {\bf r}_{j}$, $\delta_{ij} = D-|{\bf r}_{ij}|$, ${\bf n}_{ij} = {\bf r}_{ij}/|{\bf r}_{ij}|$, ${\bf v}_{n_{ij}} = [({\bf v}_{i} - {\bf v}_{j}) \cdot {\bf n_{ij}}] {\bf n}_{ij}$. The effective mass is usually $m_{eff} = M M /2 M = 1/2$ (but see below for the interaction with particles of the confining boundaries). The parameter $\gamma_n$ is fixed is such a way that the restitution coefficient is $e = 0.88$.\\
\noindent The tangential component is given by:
\[
 {\bf F_{t_{ij}}} = -k_t {\bf u}_{\bf t_{ij}} - \gamma_t m_{eff} {\bf v}_{{\bf t}_{ij}},
\]
where ${\bf u}_{\bf t_{ij}}$ is the elastic tangential displacement, and ${\bf v}_{{\bf t}_{ij}} = {\bf v}_{{ij}} - {\bf v}_{{\bf n}_{ij}}$.
${\bf u}_{\bf t_{ij}}$, set to zero at the beginning of a contact, measures the shear displacement during the lifetime of a contact. Its time evolution is fixed by ${\bf v}_{{\bf t}_{ij}}$, ${\bf \omega}_i$ and ${\bf \omega}_j$, as described in L. E. Silbert et al. {\it Phys Rev.} E, {\bf 64}, 051302 (2001). The presence of tangential forces implies the presence of torques, ${\bf \tau}_{ij} = -1/2 {\bf r}_{ij} \times {\bf F}_{t_{ij}}$. 
The shear displacement is set to zero both when a contact finish ($\delta_{ij} < 0$), and the Coulomb condition $|{\bf F_{t_{ij}}}| \leq  |\mu {\bf F_{t_{ij}}}|$ is always enforced. Here $\mu$ is the coefficient of static friction.

As described in Sec.~\ref{Sec:system}, particles are enclosed between two rough plates. Each plate is made by a collection of particles that move as a rigid object. The bottom plate is fixed, and its particles are therefore considered to have an infinite mass. The top plate has a mass equal to the sum of the masses of its particles. The masses of the confining plates enter in the calculation of the effective mass in the interaction law.

We use the value of the parameters of L. E. Silbert et al. {\it Phys Rev.} E, {\bf 64}, 051302 (2001): $k_n = 2~10^5$, $k_t/k_n = 2/7$, $\gamma_n = 50$, $\gamma_t/\gamma_n = 0$. Different values of the friction coefficient are investigated. Lengths, masses, times and stresses are measured in units of $d_0 = D$, $m_0 = M$, $t_0 = \sqrt{M/k_n}$, $\sigma_0 = k_n/D$. 

\subsection*{Preparation protocol}
The packings are constructed by first randomly placing the particles into the system with small radii, is such a way that no particles touch. Molecular Dynamics fictionless simulations are then performed by quickly inflating the particles radii in the presence of a small viscous damping force, until the radii reach their final value. Then, the system is allowed to relax until the kinetic energy vanishes. Friction is switched on after this procedure.\\
Using this procedure, the volume fraction above which only jammed (finite pressure) states are generated is $\phi_{rcp} \simeq 0.645$ [H.P. Zhang and H.A. Makse, Phys. Rev. E {\bf 72}, 011301 (2005)]. $\phi_{rcp}$ is approached exponentially fast as the system size increases  (unpublished).\\
Introducing friction after the preparation of the system allows the easy creation of dense packing of frictional systems. Experimentally, these high density states are generated via more complex procedures which allow for the continuous breaking of frictional contacts. Typical examples are vertical tapping, continuous high-frequency small amplitude vibrations, or thermal cycling.\\
The influence of the preparation protocol is described in Sec.~\ref{Sec:preparation}.

\subsection*{Computational details}
We solve the equations of motion of the system, $m \ddot{\bf r}_i = \sum_j {\bf F_{n_{ij}}} + {\bf F_{t_{ij}}}$ and $I \dot{\bf \omega}_i = \sum_j {\bf \tau }_{ij}$ via a velocity Verlet scheme, with an integration timestep $\delta t = 10^{-4}$. 
The system reaches its steady state after a time of the order of $T = 10^2$ in all regions of the phase diagram, but for the `Flow \& Jam' region, as long as $\sigma \ge 2~10^{-3}$ (the minimum value we have considered). In the `Flow \& Jam' region, the system jams after a mean time which diverges on increasing the density, and therefore the steady state can be obtained only in a small volume fraction range. In this region, we have performed simulations lasting up to a time $T = 5~10^4$. 
The simulations have been performed on a number of computer clusters. In $24h$, we simulate approximately a time $10^3$, depending on the number of particles. We have performed simulations lasting up to $50$ days.
For each considered $\phi,\sigma,\mu$ point, we have performed at least $10$ different runs, starting from different initial conditions. In the ``Flow \& Jam region'', where we need statistics to properly measure the mean jamming time, we have performed $100$ runs for each considered $\phi,\sigma,\mu$ point.

\section{Investigated system~\label{Sec:system}}
We have investigated via Molecular Dynamics simulations a system of grains confined between two rough plates, as illustrated in Figure S1. The vertical distance between the plates is fixed, and a shear stress $\sigma$ along $x$ is applied to the top plate.

\begin{figure}[h!!!]
\begin{picture}(200,60)(0,0)
\put(40,0){\includegraphics*[scale=0.25]{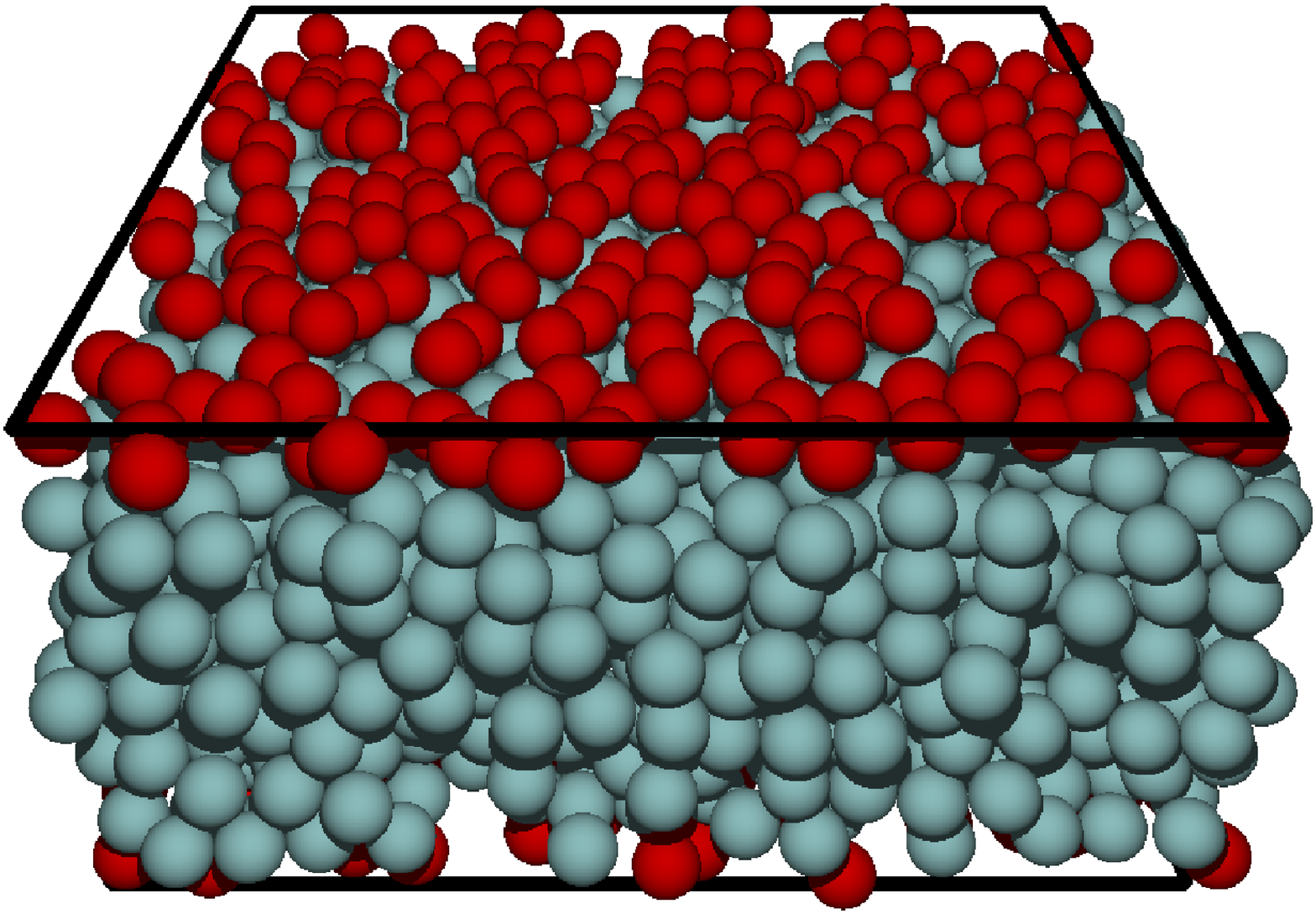}}
\end{picture}
\end{figure}
{\bf Figure S1:} the investigated system. Grains are confined between two rough plates (red particles) at a fixed vertical distance. A shear stress directed along $x$ is applied to the top plate, while the bottom one is kept fixed.

\subsection*{System size}
Particles are enclosed in a box of dimension $l_x = l_y = 16D$, and $l_z = 8D$. Periodic boundary conditions are used along $x$ and $y$. The size of the vertical dimension $l_z$ is chosen to be comparbale to that of recent experiments [D. J. Pine, contribution to KITP Program on Granular Physics, 2005, unpublished; J.-C. Tsai and J. P. Gollub, {\it Phys. Rev.} E {\bf 72}, 051304 (2005); K. E. Daniels and R. P. Behringer, {\it Phys. Rev. Lett.} {\bf 94} 168001 (2005)]. We have investigated the effect of the finite size of our system investigating system with $l_z$ up to $64D$. The effect of the system size is described in Sec.~\ref{SecSize}.

\subsection*{Volume fraction}
The volume fraction $\phi$ is equal to the volume occupied by the grains divided by the volume of the container, i.e. $\phi = Nv_0/V_0$, where $V_0 = l_x l_y l_z$ is the volume of the system, $N$ is the number of enclosed grains, and $v_0= 1/6 \pi D^3$ is the volume occupied by a single grain. 

Here, we have defined the volume fraction introducing a term which takes into account the effect of the rough plates protruding into the system. Due to the boundaries, the volume accessible to the grains is not $V_0$, but $V = V_0 -\Delta V$, where $\Delta V$ is an unknown corrective term. Since $\Delta V$ is much smaller than $V_0$, we have:
\[\begin{array}{lcr}
{\rm \hspace{3cm}}
\phi(N) = \frac{N v_0}{V_0 - \Delta V} \simeq \frac{N v_0}{V_0} \left(1-\frac{\Delta V}{V_0}\right) & 
{\rm \hspace{2.5cm}Eq.~S1}
\end{array}
\]
In order to estimate $\Delta V$ we have determined the maximum number of grains $N_{max}$ which is possible to enclose into the system at zero applied stress, using the protocol described above to prepare the initial packing. We have then fixed $\phi(N_{max}) = \phi_{rcp}$ and determined $\Delta V$ inverting Eq.~S1.

\newpage
\section{Dynamical regimes}
Fig. S2 shows in the upper (lower) panel the time evolution of the position (velocity) in the Flow, Flow \& Jam, Slip \& Jam and Jam (in the insets) regions for $\sigma = 2~10^{-3}$. The response of the system in the different regions can be summarized as follows:\\
\noindent {\bf Flow:} the system flows with a steady velocity reached after a transient.\\
\noindent {\bf Flow \& Jam:} the system reaches a steady velocity after a transient. However, after flowing for sometime, it suddenly jams.\\
\noindent {\bf Slip \& Jam:} steady flow is never observed. The system jams after a small inelastic displacement of the top plate. \\
 \noindent {\bf Jam:} the system responds as an elastic solid to the applied stress.

\begin{figure}[!h]
\begin{center}
\includegraphics*[scale=0.33]{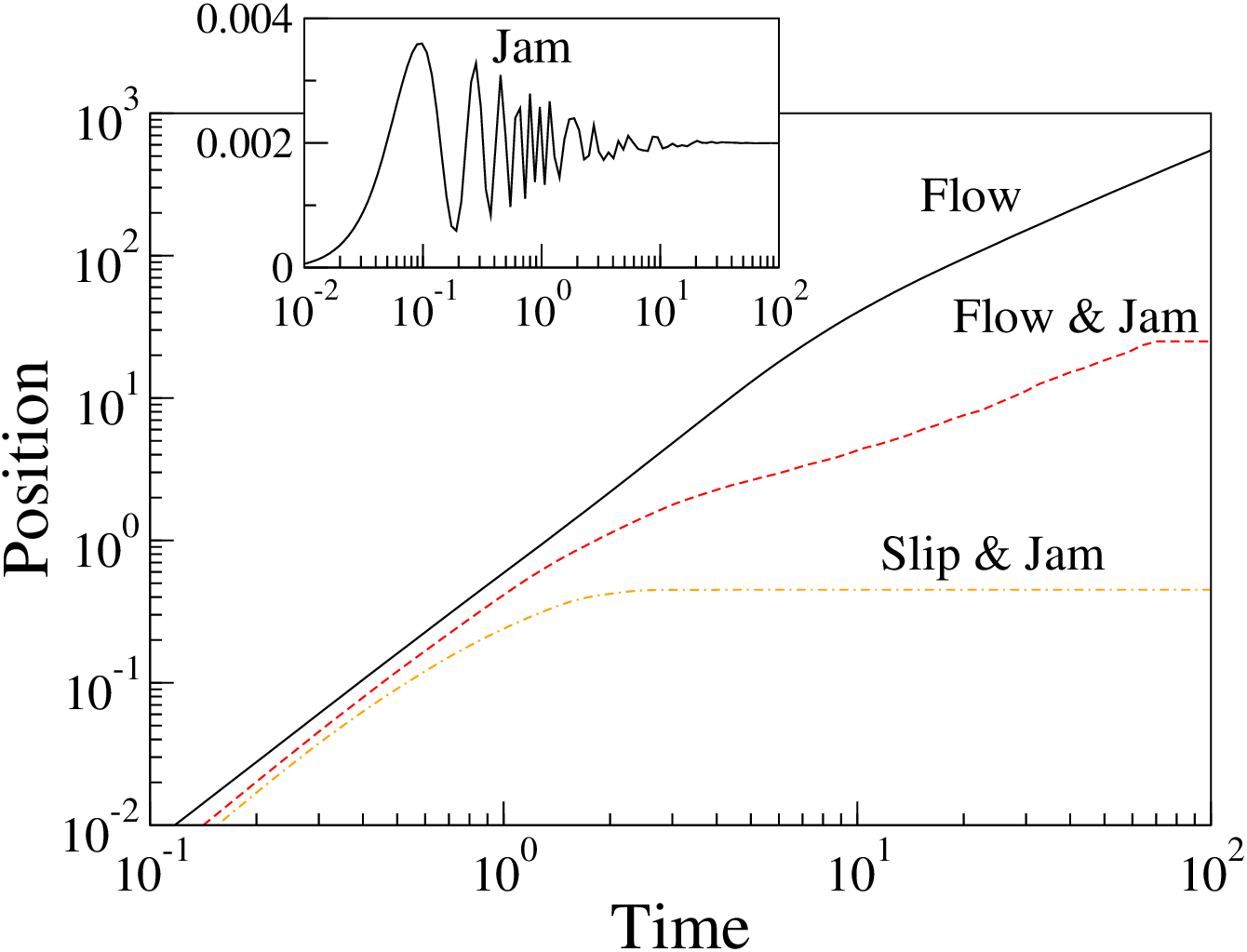}\\
\noindent \includegraphics*[scale=0.33]{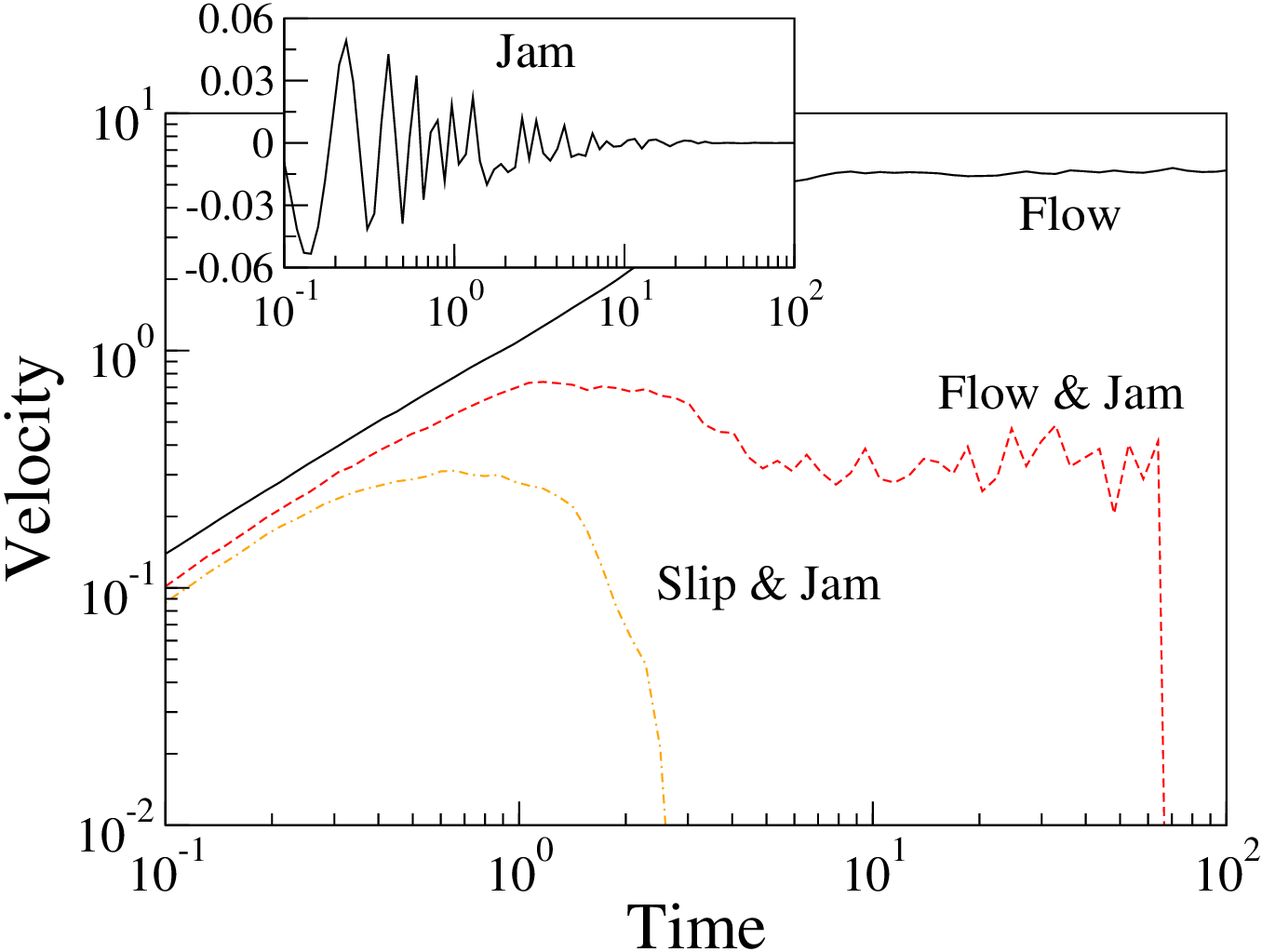}
\end{center}
{\bf Figure S2:} Dynamical regimes of a sheared granular system. 
\end{figure}

\newpage
\section{Definition of $\pc$}
We have defined the slip distance $\Delta L (\phi,\sigma)$ as the displacement of the top plate in a stress cycle. After preparing the system, we slowly increase the shear stress to its final value $\sigma$, and then decrease it to zero. Figure S3 (top panel) shows the displacement of the top plate position as a function of the shear stress for $\sigma = 2~10^{-3}$ and $\mu = 0.1$. Different curves refer to different values of the volume fraction, as shown. At small $\phi$, the initial and final position of the top plate do not coincide: the residual displacement is our measure of the slip, $\Delta L$, as exemplified in the figure for $\phi = 0.642$. The bottom panel shows that the slip distance decreases as a power law as the volume fraction increases: this allow to define $\pc(\sigma,\mu)$ as the volume fraction where $\Delta L$ vanishes.

\begin{figure}[h!]
\begin{center}
\begin{picture}(400,42)(0,0)
\put(15,0){\includegraphics*[scale=0.25]{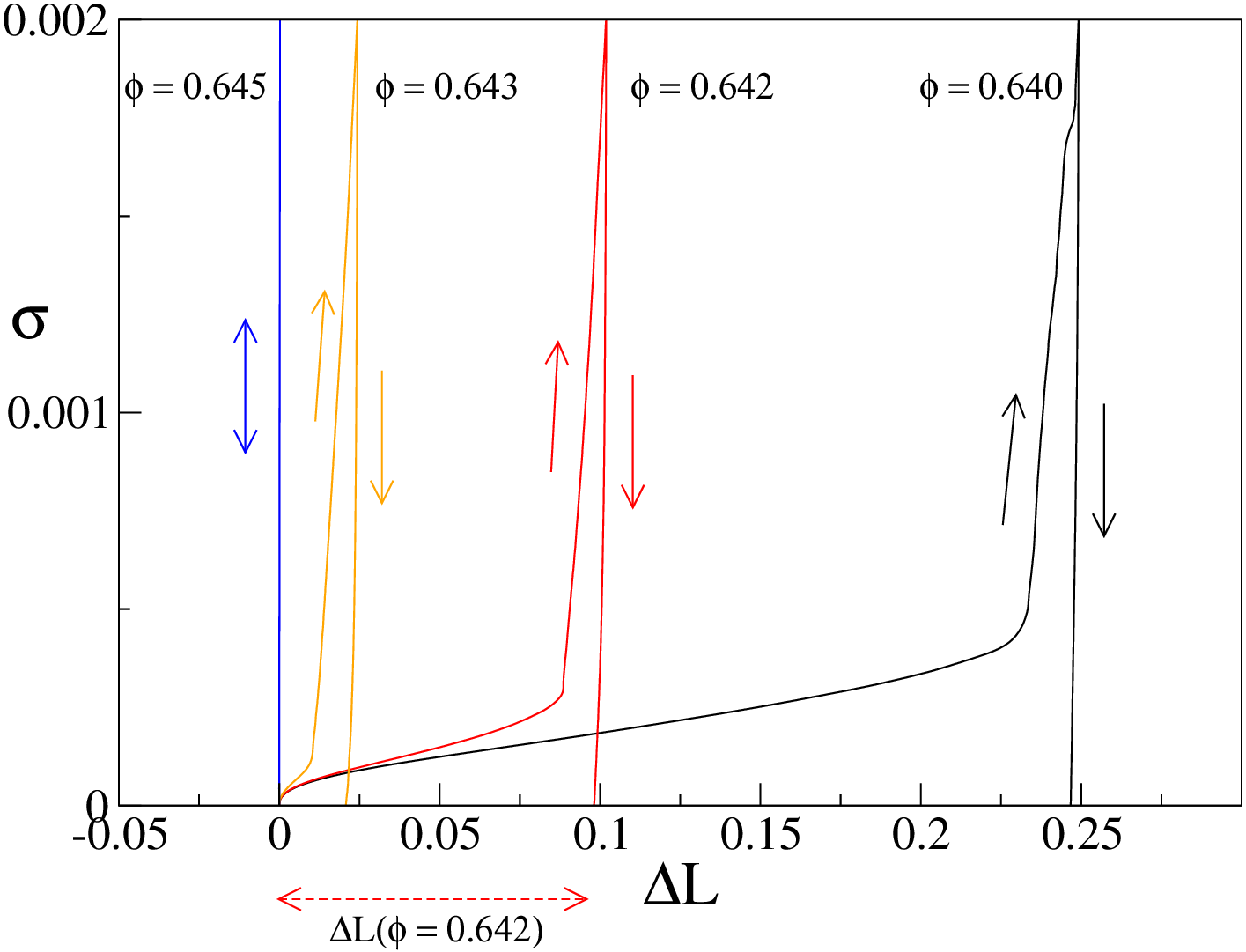}}
\put(80,0){\includegraphics*[scale=0.24]{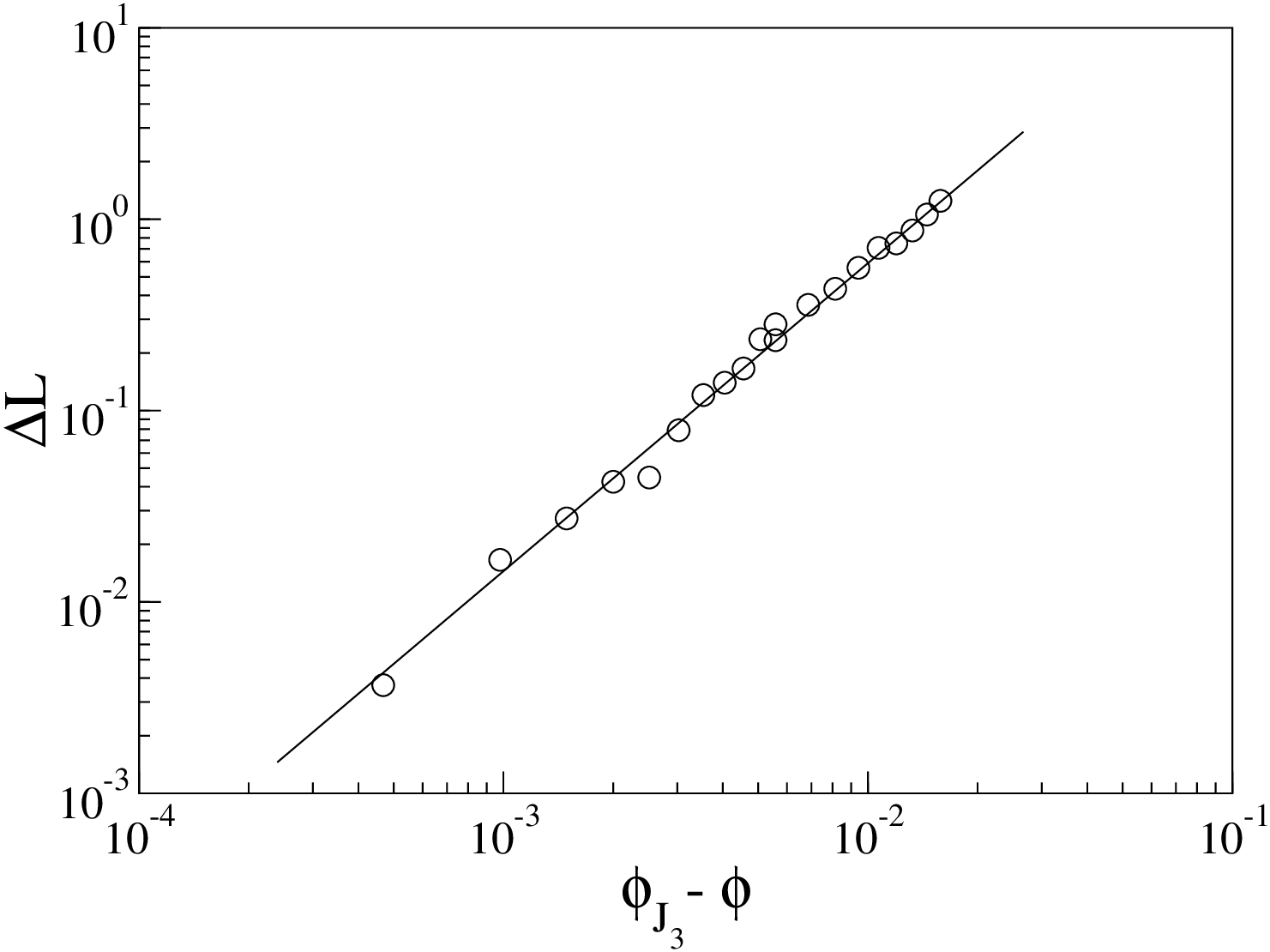}}
\end{picture}
\end{center}
\end{figure}
{\bf Figure S3:} Left: displacement of the top plate in a stress cycle. The stress is first increased to its final value $\sigma$, and then decreased to zero. The residual displacement is our definition of the slip $\Delta L$. Right: for a fixed value of $\sigma$, the slip decreases on increasing the volume fraction, and vanishes at a volume fraction $\pc$, which depends on $\sigma$ and $\mu$. The straight line is a power law $\Delta L = a (\phi-\pc)^b$, with $a \simeq 10^3$, $b \simeq 1.6$.\\
\\
\noindent At very high values of the shear stress it is not possible to define $\pc$ this way, as one finds $\Delta L > 0$ even at very high volume fractions. The reason is that at high shear stress the system behaves plastically: one finds $\Delta L > 0$ not because the system slips, but because it deforms plastically in the stress cycle. However, when this is the case the dependence of $\Delta L$ on $\phi$ shows a clear crossover from a slip-dominated regime to a plastic-dominated regime, as shown in Fig. S4. This crossover allows to define $\pc$ as the inflection point of $\Delta L(\phi)$.

The crossover from the elastic to the plastic regime is due to the increase of the number of contacts that are broken as the strain increase. At small $\sigma$, the strain of the system is small, and contacts do not break. At higher $\sigma$, the strain of the system is large, and contacts break. When a contact breaks, the tangential force between the grains is irreversibly destroyed, which is the microscopic origin of the plastic response of the system.

\begin{figure}[h!]
\begin{center}
\includegraphics*[scale=0.22]{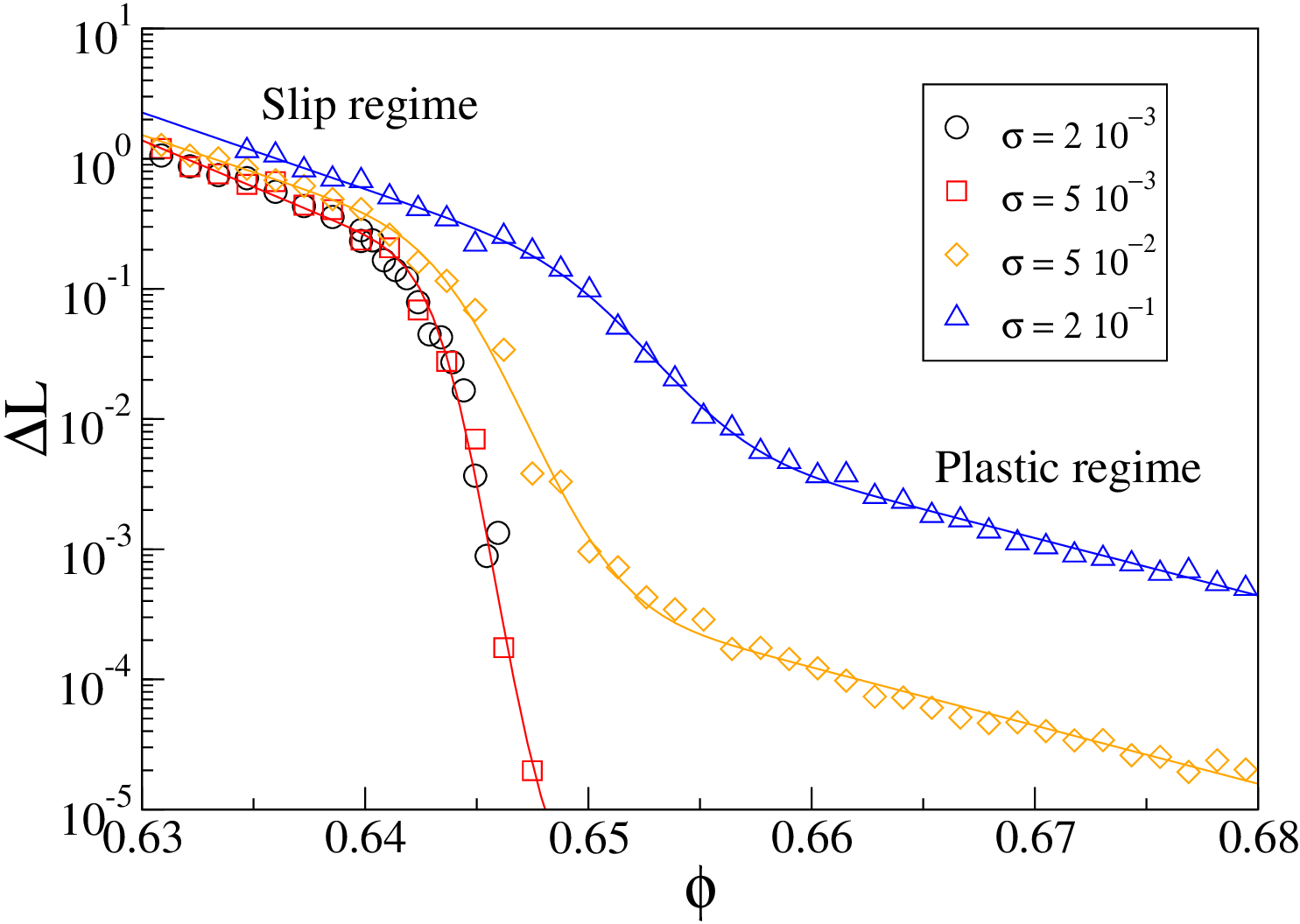}
\end{center}
\end{figure}
{\bf Figure S4:} At small $\sigma$ $\Delta L$ vanishes at high $\phi$, and $\pc$ is defined as the volume fraction where $\Delta L$ vanishes as illustrated in Fig.S3. On the contrary at high $\sigma$, $\Delta L$ does not vanishes, but shows a crossover from a a slip-dominated regime to a plastic-dominated regime. In this case $\pc$ is defined as the inflection point of $\Delta L(\phi)$. 

\newpage
\section{Measure of the shear modulus G}
To measure the shear modulus $G$ we have applied to a system jammed under the action of a shear stress $\sigma$ a perturbing shear stress $\delta \sigma$. The only non-zero components of $\delta \sigma$ are $\delta \sigma_{xx}$ and $\delta \sigma_{yy}$, fixed such as $\delta \sigma_{xx}^2 + \delta \sigma_{yy}^2 = \delta\sigma^2$. The perturbing shear stress is therefore conveniently expressed in terms of its magnitude $\delta \sigma$ and of $\theta = \arctan\left(\delta \sigma_{yy}/\delta \sigma_{xx}\right)$.
The shear modulus $G$ is defined as $\lim_{\delta \sigma \to 0} \delta \sigma/\epsilon$, where $\epsilon$ is the shear strain induced by $\delta \sigma$.
This definition is appropriate as for small $\delta\sigma$ ($\delta\sigma < 10^{-3}\sigma$) the response of a jammed system is elastic (the strain is proportional to the stress) and to a good approximation isotropic ($\epsilon$ only very weakly depend on the direction of $\delta\sigma$ with respect to that of $\sigma$, see below). 
Fig. S5 show the displacement $\delta(r) = (\delta x,\delta y)$ of the top plate position for different values of the volume fraction ($\phi$ varies from $\phi = 0.626 $ to $\phi = 0.663$, the smallest $\phi$ corresponding to the largest circle).
Each curve is obtained by first applying a perturbing shear stress with ($\theta = 0$), and then by increasing $\theta$ from $0$ to $2\pi$. The figure clarifies that systems jammed under shear are elastic, as each curve describes a close path.

This result also clarifies that, even though the mechanical rigidity of a system jammed under shear stress originates from an underlying force network which is highly anisotropic, and which builds up only because of the presence of an applied shear stress, yet the system behaves as an elastic solid when a small perturbing shear stress is applied. It is therefore not `fragile' as recently speculated [M.E. Cates, J.P. Wittmer, J.P. Bouchaud, P.Claudin, Phys.~Rev.~Lett. {\bf 81}, 1841-1844 (1998)]. We have checked that a fragile behaviour shows up in the response to larger perturbations.

\begin{figure}[!ht]
\begin{center}
\begin{picture}(300,60)(0,0)
\put(50,0){\includegraphics*[scale=0.33]{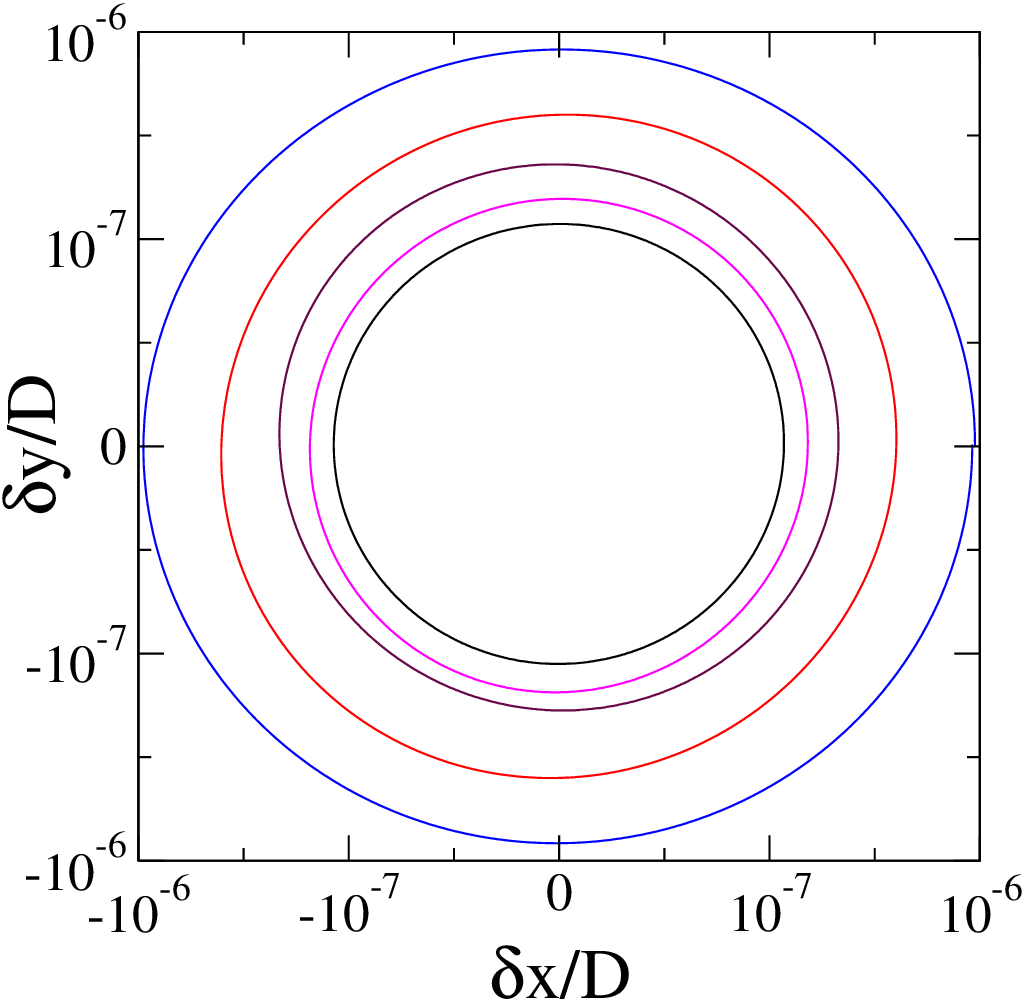}}
\put(0,-5){{\bf Figure S5:} Response of jammed system to a small perturbing shear stress.}
\end{picture}
\end{center}
\end{figure}

The data of Fig.S5 suggests that the system behaves isotropically. To check whether this is actually the case, we have investigated the parameter 
\begin{equation}
 \xi(\theta) = \frac{\left[ \delta x^2 (\theta) + \delta y^2 (\theta) \right]^{1/2} - \overline{\delta r} }{\overline{\delta r}},
\end{equation}
where $\overline{\delta r} = \langle \left[ \delta x^2 (\theta) + \delta y^2 (\theta) \right]^{1/2} \rangle_\theta$. $\xi(\theta)$ measures how close to the mean behavior the system is at each value of $\theta$. Figure S6 shows that $|\xi(\theta)| < 4\%$, suggesting that the response of the system is isotropic to a very good approximation.
\pagebreak
\begin{figure}[!!h]
\begin{center}
\includegraphics*[scale=0.25]{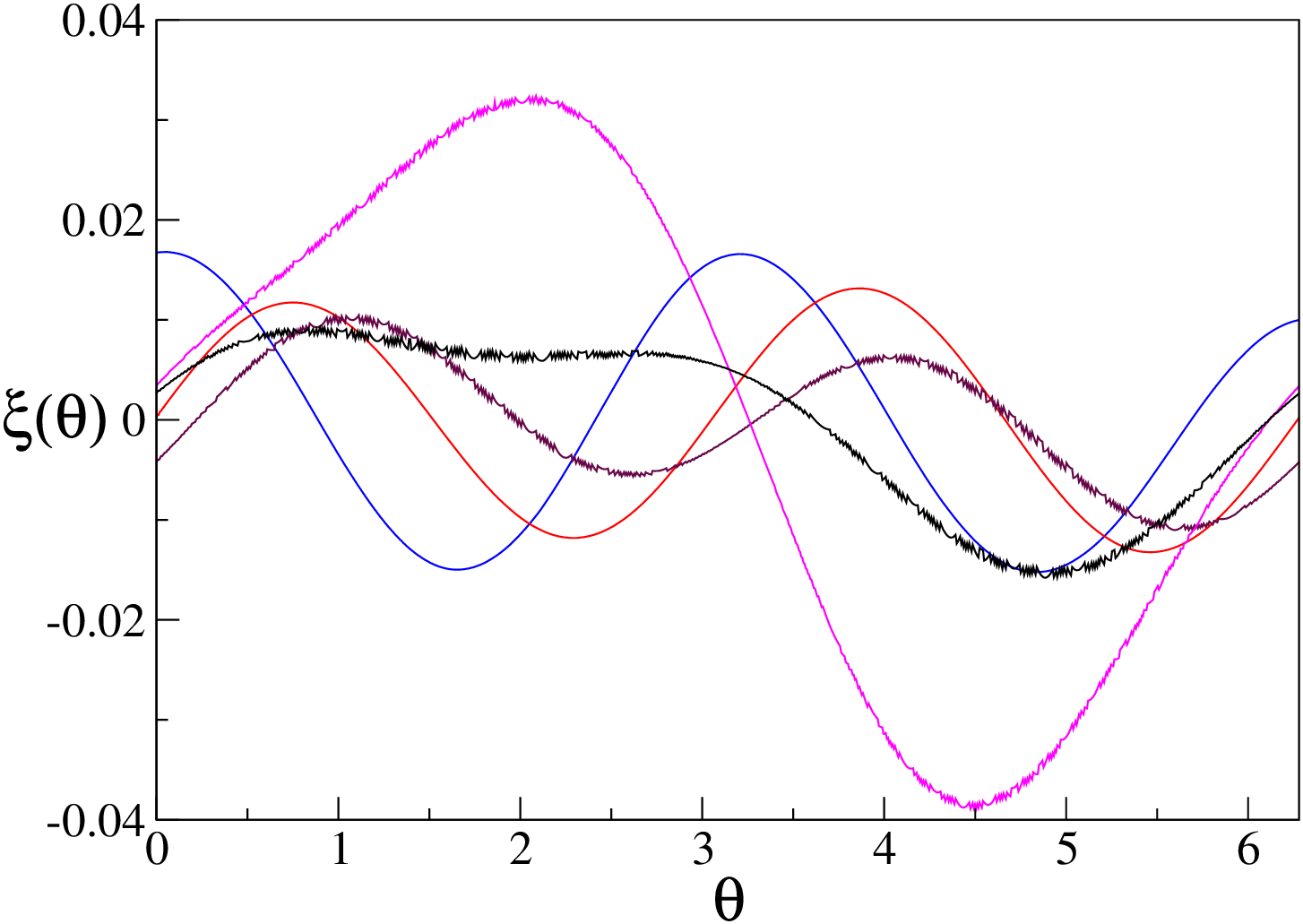} 
\end{center}
\end{figure}\\
\noindent {\bf Figure S6:} Anysotropy in the response of jammed system to a small perturbing shear stress. Different curves refer to different values of the volume fraction.\\

\newpage
\section{Finite-size analysis\label{SecSize}}
In this section, we discuss the robustness of the jamming phase diagram for frictional particles described in the main text to variations of the system size. We have kept fixed the size of the system in the transverse directions, $l_x = l_y = 16D$, and varied the vertical size $l_z$. We compare the results for $l_z = 8$, which are the ones described in the manuscript, with results obtained with $l_z = 16$ and $l_z = 32$, as obtained for $\sigma = 2~10^{-2}$ and $\mu = 0.1$.

\subsection*{Finite size effects at $\phi_{J_3}$}
At the jamming line $\phi_{J_3}$, defined as the volume fraction at which the `slip' vanishes, structural quantities have cusps, as shown in Fig.~3 of the manuscript. To investigate the dependence of the line $\phi_{J_3}$ on the system size, we have studied the size dependence of the location of the cusp in the pressure on the system size. As shown in Fig.~S7, the cusp always occurs at the same volume fraction, implying that the line $\phi_{J_3}$ does not depend on the size of the system. This is not a surprise because (at small $\sigma$) $\phi_{J_3}$ coincides with the random close packing volume fraction, which approaches exponentially fast its asymptotic value with the size of the system (when packings are generated with the protocol described in Sec.~\ref{Sec:numericalmodel}).

\begin{figure}[!!h]
\begin{center}
\includegraphics*[scale=0.33]{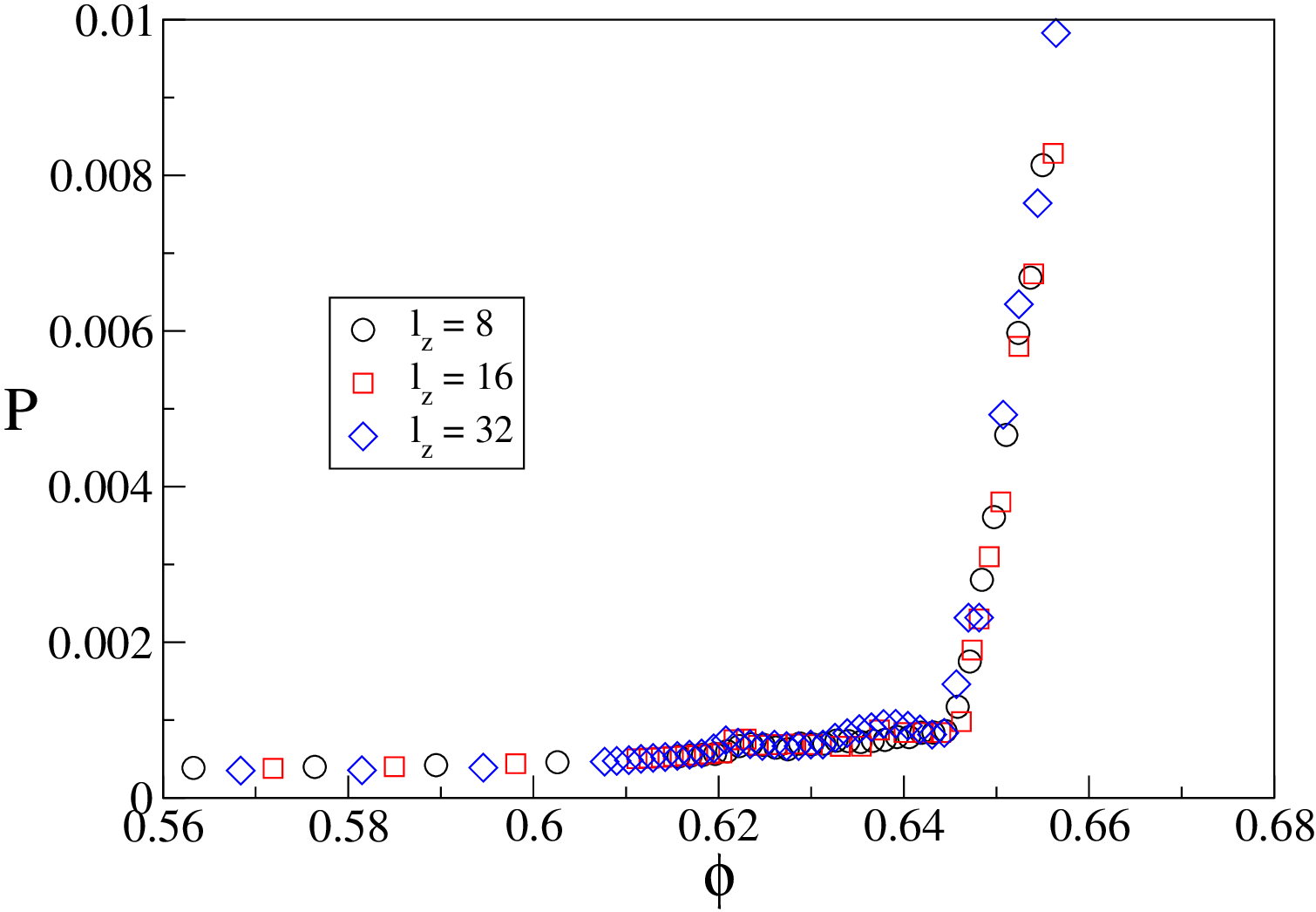} 
\end{center}
\end{figure}
{\bf Figure S7:} Normal pressure acting on the top confining plate as a function of the volume fraction, for different system sizes. 

\newpage
\subsection*{Finite size effects at $\phi_{J_2}$}
For each value of $l_z$, we have measured the shear viscosity $\eta$ in the steady state, which appears to diverge as a power-law as the density increases. As shown in Fig.~S8, data obtained with different sizes can be reasonably scaled on the same curve indicating that our system so large enough that finite size effects are small. The number of particles at $\phi_{J_2}$ varies between $2400$ and $9600$, depending on the size 

As $\phi$ approaches $\phi_{J_2}$, one enters the `Flow \& Jam' region of the phase diagram, where the system jams after flowing in a steady state for a time $t_{\rm jam}$. Since $t_{\rm jam}$ becomes smaller and smaller as $\phi$ approaches $\phi_{J_2}$, it is not possible to obtain reliable steady state shear viscosity data very close to $\phi_{J_2}$. 

\begin{figure}[!!h]
\begin{center}
\includegraphics*[scale=0.3]{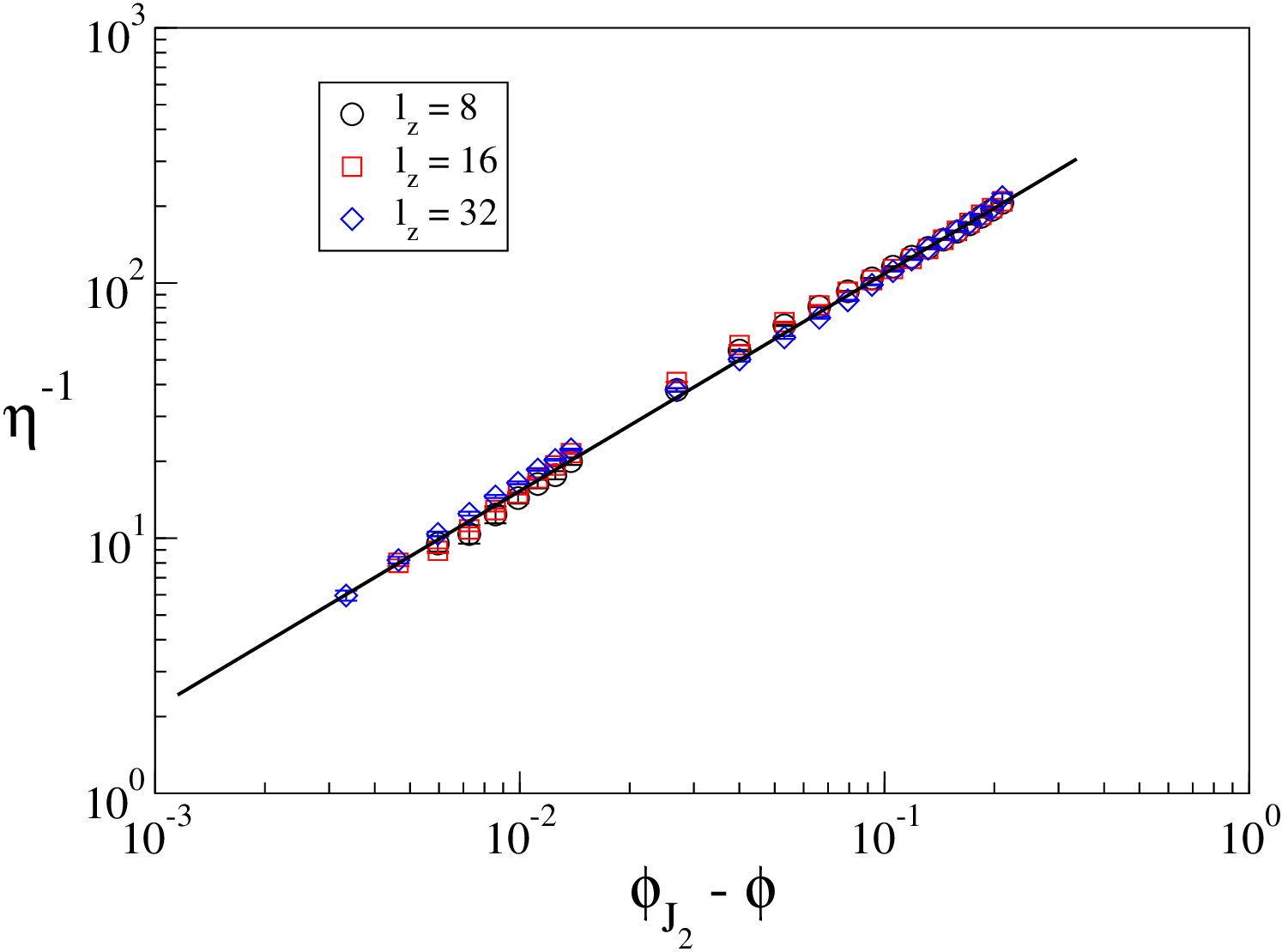} 
\end{center}
\end{figure}
{\bf Figure S8:} Log-log plot of the inverse shear viscosity $\eta^{-1}$ versus $\phi_{J_2}-\phi$, for different system sizes. The data collapse on the same master curve ($\eta^{-1} \simeq (\phi_{J_2}-\phi)^\gamma$, $\gamma \simeq 0.85$), indicating that 
finite-size effects are negligible.\\

\newpage
\subsection*{Finite size effects at $\phi_{J_1}$}
The jamming volume fraction $\phi_{J_1}$ is that where the time $t_{\rm jam}$ a system flows in a steady state before jamming diverges on decreasing the volume fraction. Since its definition involves a diverging time scale, its numerical identification is difficult, as well as the understanding of its dependence on the size of the system. To check for the presence of finite-size effects we have computed, for any given value of $l_z$, the probability $p$ that a simulation jam in a given time $T$ as a function of $\phi$. The probability is computed over $100$ runs which differ for the initial conditions, while the simulation time is fixed to $T = 100$. The results are shown in Fig. S9.

\begin{figure}[!h]
\begin{center}
\includegraphics*[scale=0.3]{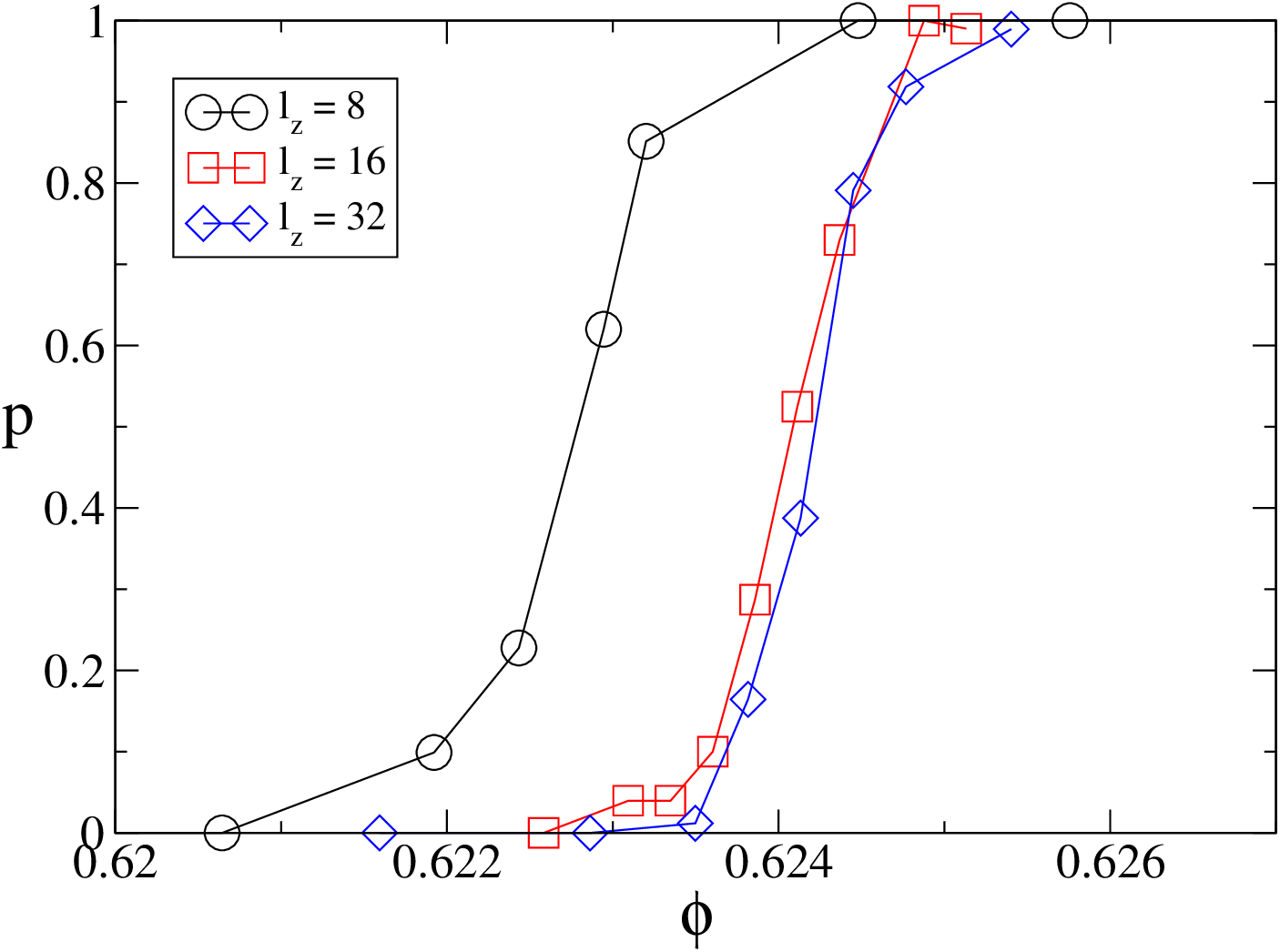} 
\end{center}
\end{figure}
{\bf Figure S9:} Fraction of simulation (over $100$) that jams in a time $T = 100$ as a function of the volume fraction, for different system sizes.\\
\\
\noindent As the system size increases, at any given value of $\phi$ the fraction of runs which jam in a time $T$ decreases. However, considering that we have only investigated a finite time $T$, and that the time required for a system to jam is expected to grow with the system size, one cannot draw from Fig.~S9 any conclusion regarding the behavior of the line $\phi_{J_3}$ in the infinite system size, infinite $T$ limit. Nevertheless, the flow \& jam phenomenology appears to be relevant, as we observe it in systems with a size comparable to that of many granular experiments. We expect the phenolenology to be actually more apparent in experiments, where one can investigate a time $T$ much larger than the one accessible in molecular dynamics simulations.

Finally, we note that the flow \& jam phenomenology is possibly related to the giant stress fluctuations observed in granular systems sheared at constant rate and constat volume fraction [see, for instance, B. Miller, C. O'Hern, and R. P. Behringer, Phys. Rev. Lett. {\bf 77}, 3110 (1996)]. At constant shear stress we do not observe large stress fluctuations, but large velocity fluctuations, the system jamming when there is a fluctuation so large that the velocity vanishes.
In both cases the large fluctuations are observed in systems large enough to be of practical interest, while it is not clear if the fluctuations persist in the infinite system size limit.

\newpage
\section{Influences of the preparation protocol~\label{Sec:preparation}}
Due to the presence of frictional forces, the response of granular systems to applied perturbations may depend on the particular protocol used to prepare the initial state. The phenomenology we have presented has been observed using a preparation protocol in which frictional forces are introduced after the system has reached a state of zero kinetic energy at the desired volume fraction. Our initial state is therefore memoryless. This protocol allows to access the whole zero pressure jamming phase diagram [C. Song, P. Wang and  H.A. Makse, Nature {\bf 453}, 629 (2008).]
Experimentally, the expectation is that our initial states are those obtained compacting a granular system, for instance via high-frequency small-amplitude vibrations, which are able to destroy frictional contacts [G-J Gao, J. Blawzdziewicz, CS. O'Hern and M. Shattuck arXiv:0907.2106 (2009)], or via similar procedures. 

Here, we consider how our findings change when the initial packing is prepared using a different and popular protocol [H. P. Zhang and H. A. Makse,  Phys. Rev. E {\bf 72}, 011301 (2005); E. Somfai, M. van Hecke. W.G. Ellenbroek, K. Shundyak, W. van Saarloos, Phys. Rev. E {\bf 75} 020301 (R) (2007); and many others], where friction is always taken into account. Grains, initially placed in random positions with small radii, are inflated until they reach their final size. During inflation, frictional contacts are taken into account. We use the same inflation rate $\Gamma$ both when using the protocol considered in the manuscript  (`no friction protocol'), as well as when using the modified protocol (`friction protocol'). In Fig.~S10 we compare, for $\sigma = 2~10^{-3}$ and $\mu = 0.1$, the velocity of the shear plate (upper panel), and the pressure (lower panel) as obtained using the two protocols. The pressure is normal force acting on the top plate divided by its surface. 

The shear velocity is the same regardless of the initial protocol, in agreement with the expectation that the flowing systems don't remember their initial state. Accordingly, the line $\phi_{J_2}$, where the viscosity diverges (the velocity vanishes), is protocol-independent. The same is true for the line $\phi_{J_1}$ (not show), which is determined from the divergence of the jamming time, also measured when the system flows.

The pressure, which is shown in the bottom panel, has a cusp at $\phi_{J_3}$ (at small $\sigma$). Figure S10, therefore, clarifies that the line $\phi_{J_3}$ depends on the preparation protocol. However, the line obtained with the `no friction' protocol used in this work is a special one, and in this sense has to be preferred over the others, as it is an upper bound with respect to all possible lines obtained using different preparation protocols. 

\begin{figure}[!!h]
\begin{center}
\includegraphics*[scale=0.33]{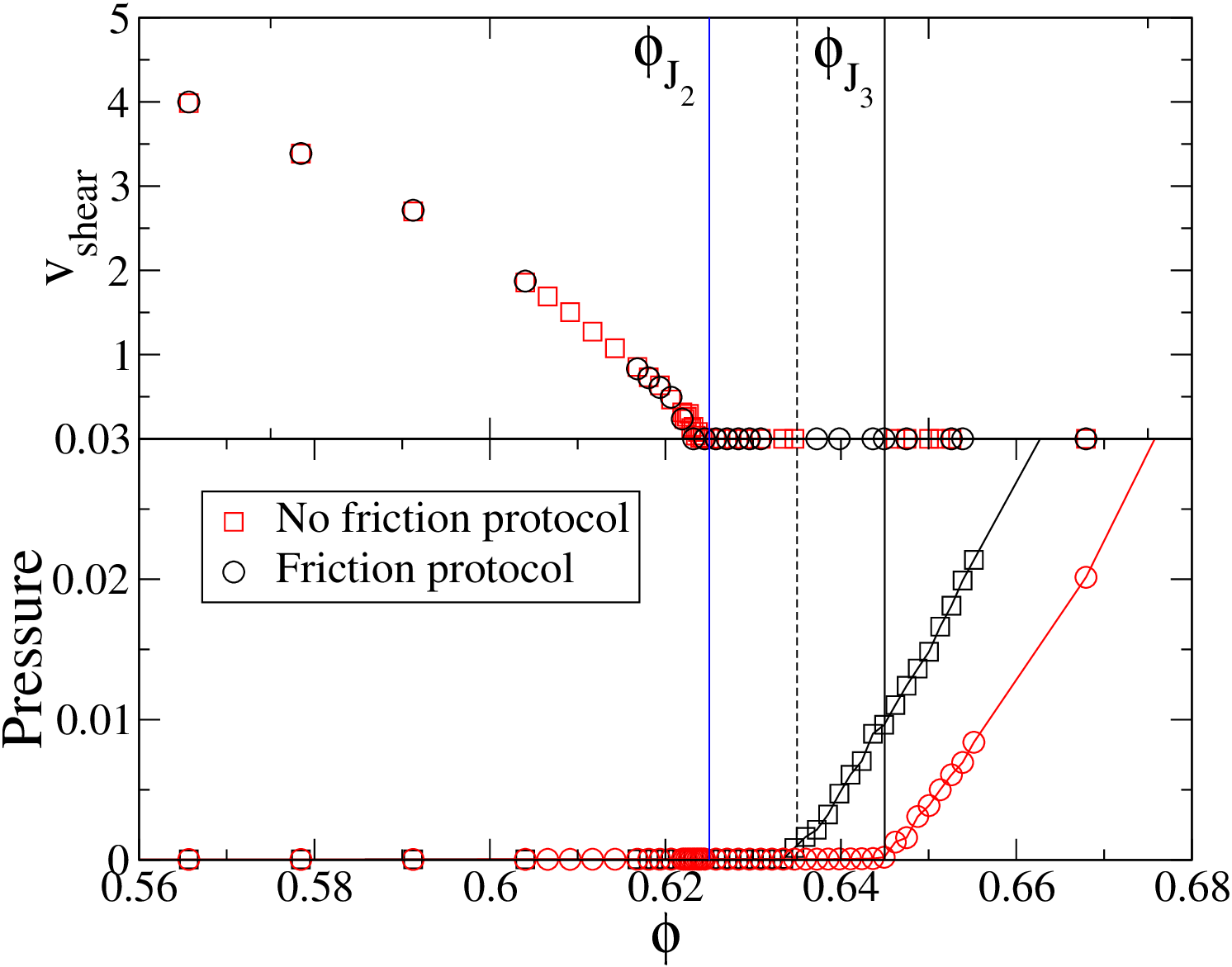} 
\end{center}
\end{figure}
{\bf Figure S10:} Location of the jamming transition  lines using different protocol to prepare the initial state. $\phi_{J_1}$ and $\phi_{J_2}$ are protocol independent, while the line $\phi_{J_3}$ depends on the protocol. Our estimate for $\phi_{J_3}$ is an upper bound for all possible estimations one can obtain using different protocols.

\newpage
\section{Dynamics of jamming~\label{Sec:dynamicsjamming}}
In the `Flow \& Jam' region of the jamming phase diagram, a system flows for a time $t_{jam}(\phi)$, but then suddenly jams. This is a peculiar fluid to solid transition since it is not driven by changes in the control parameters - it is the result of a fluctuation which brings a flowing system in a jammed configuration able to support the applied stress.

In Fig.~S11, we show the time evolution of several quantities as the system flows and then jams, at $\phi \simeq 0.622$, $\sigma = 2~10^{-3}$, and $\mu = 0.1$.
In panel (a), we plot the velocity of the top plate, which reaches a (fluctuating) steady state after a transient, but vanishes after a long time. Panel (b) shows the time evolution of the translational kinetic energy ($K_t = \sum_{i=1}^N = 1/2 m |{\bf v}_i|^2$), of the rotational kinetic energy ($K_r = \sum_{i=1}^N = 1/2 I {\bf \omega}_i^2$), of the elastic energy due the normal ($U_n = \sum_{i\neq j} = 1/2 k_0 |{\bf \delta}_{ij}|^2$) and to the tangential ($U_t = \sum_{i\neq j} = 1/2 k_t |{\bf u}_{ij}|^2$) interaction. The ratio between $U_n/U_t$ is roughly $100$, as expected as the tangential force is bounded by the normal one, $|{\bf f}_t| \leq |\mu {\bf f}_n|$, so that $U_n/U_t \simeq |{\bf f}_t|^2 / |{\bf f}_n|^2 \simeq 1/\mu^2$. When the system jams, the elastic energies reach a plateaux, while the kinetic energies vanish.
Panel (c) shows the evolution of the mean contact number, which increases on jamming. Finally, panel (d) shows the evolution of the normal pressure on the shearing plate. The pressure at each instant is equal to the force acting on the top plate divided by its surface (we are not averaging the signal over time). The pressure fluctuates about a consant value: its fluctuations are much smaller than that found at constant volume and constant shear rate [B. Miller, C. O'Hern, and R. P. Behringer, Phys. Rev. Lett. {\bf 77}, 3110 (1996)].

\begin{figure}[!!h]
\begin{center}
\includegraphics*[scale=0.35]{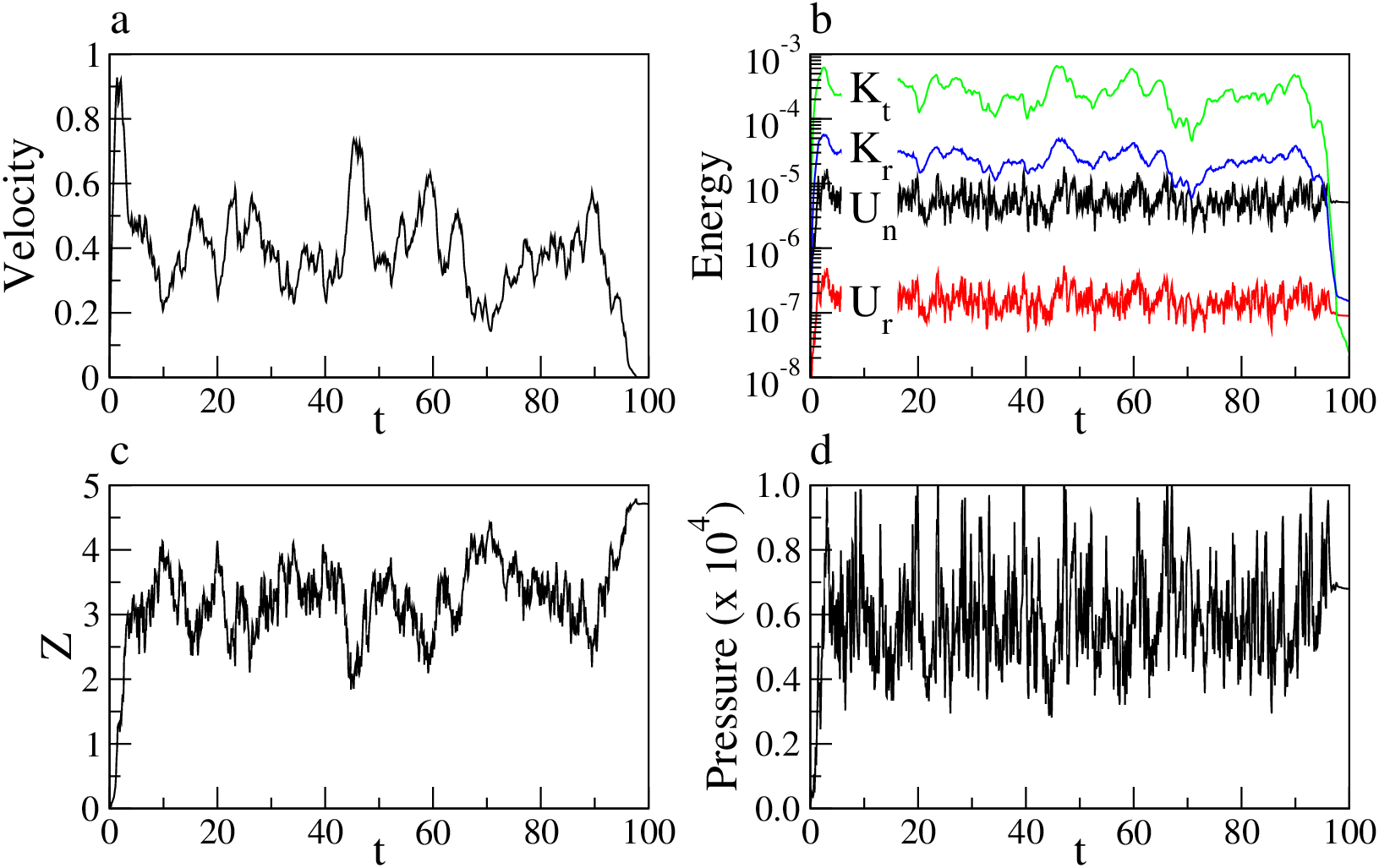} 
\end{center}
{\bf Figure S11:} A typical run in the `Flow \& Jam' region of the jamming phase diagram.
\end{figure}

\newpage
The analysis of these quantities, as well as that of the fraction of the sliding contacts (a contact slides when the Coulomb condition is enforced), shown in Fig.S12, clarifies that in the jamming transition under shear the system quickly transients from a highly dynamic regime, in which the kinetic energy is serval orders of magnitude greater than the potential energy, and almost all contacts are sliding,  to a jammed one, where the kinetic energy and the fraction of sliding contacts vanish.

\begin{figure}[!!h]
\begin{center}
\includegraphics*[scale=0.25]{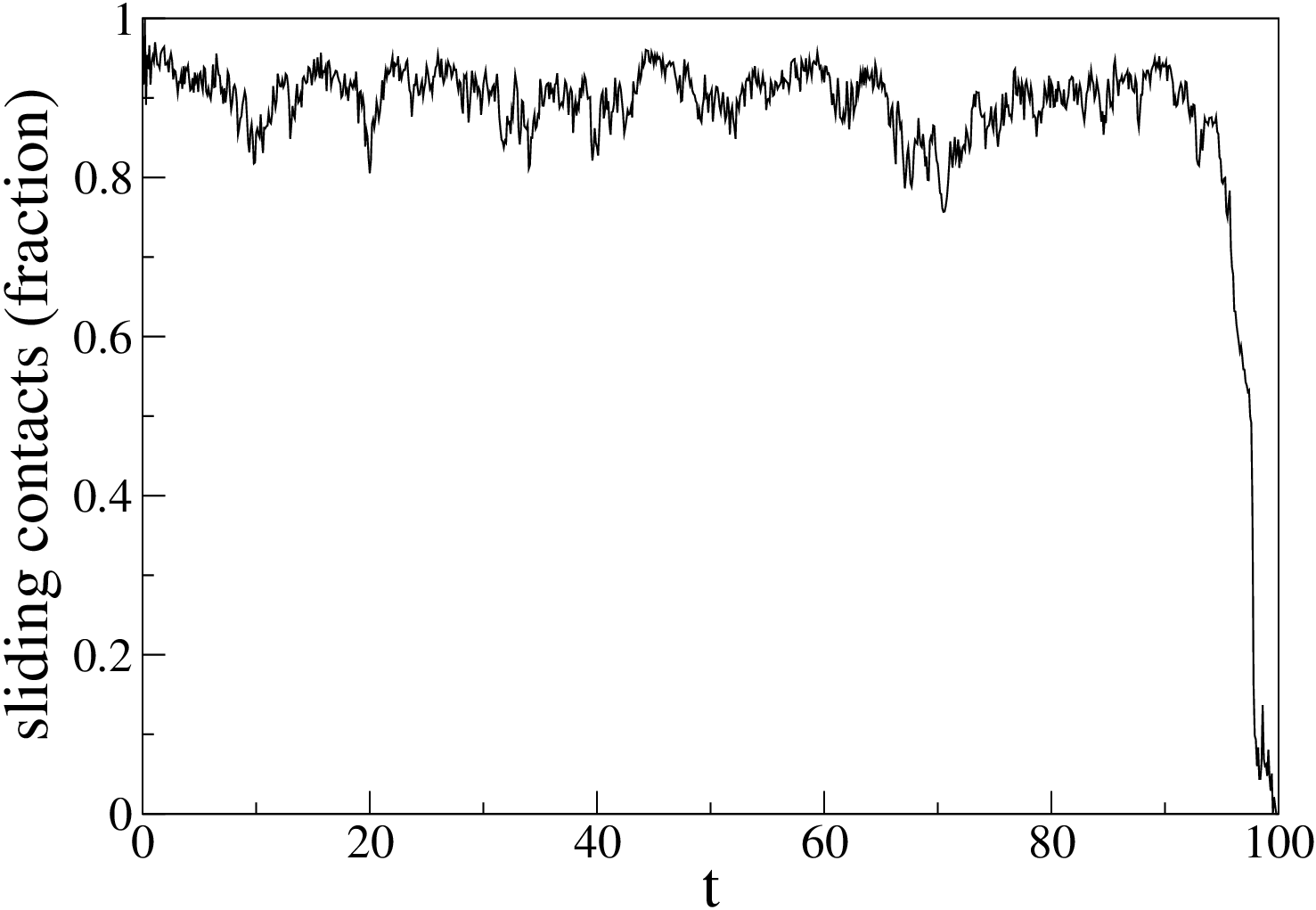} 
\end{center}
{\bf Figure S12:} Time evolution of the fraction of sliding contacts.
\end{figure}

\end{widetext}


\begin{thebibliography}{50}
\bibitem{note}
The shear velocity $v_{s}$ is the velocity of the top plate, which in the `Flow' regime is the stationary value, and in the `Flow \& Jam' region is the stationary value before jamming.
\end{thebibliography}
\end{document}